\documentclass{llncs}
\usepackage[,frozencache]{minted}
\usepackage[utf8]{inputenc}
\usepackage[T1]{fontenc}
\usepackage{xpatch,letltxmacro}
\usepackage{newunicodechar}
\usepackage{lmodern}
\usepackage{stmaryrd}
\usepackage{todo}
\usepackage{url}
\usepackage{hyperref}
\usepackage{breakurl}
\usepackage{tikz}
\usepackage{calc}
\usepackage{amssymb}
\usepackage{array}
\usepackage{wrapfig}
\usepackage[inline]{enumitem}

\newcommand{\shortdots}{.\!.\!.}

\newlist{enumi}{enumerate*}{1}
\setlist*[enumi,1]{%
  label=(\roman*),
}

\usepackage{newunicodechar}

\newcommand{\Arrow}[1]{%
     \parbox{#1}{\tikz{\draw[->](0,0)--(#1,0);}}
}
\newcommand{\printtowidth}[2]{\makebox[\widthof{#1}]{#2}}
\newcommand{\onecharwidth}[1]{\printtowidth{x}{#1}}
\newcommand{\twocharwidth}[1]{\printtowidth{xx}{#1}}
\newcommand{\threecharwidth}[1]{\printtowidth{xxx}{#1}}

\newcommand{\notPropEq}{\twocharwidth{\rlap{\hspace{1.1mm}/}\texttt{=\hspace{0.3mm}=}}}

\newunicodechar{∈}{\onecharwidth{\ensuremath{\in}}}
\newunicodechar{⊆}{\onecharwidth{$\subseteq$}}
\newunicodechar{⊑}{\onecharwidth{\scalebox{0.8}[0.95]{$\sqsubseteq$}}}
\newunicodechar{∪}{\onecharwidth{$\cup$}}
\newunicodechar{∩}{\onecharwidth{$\cap$}}
\newunicodechar{⊔}{\onecharwidth{$\sqcup$}}
\newunicodechar{⊓}{\onecharwidth{$\sqcap$}}
\newunicodechar{γ}{\onecharwidth{$\gamma$}}
\newunicodechar{→}{\onecharwidth{~\Arrow{.18cm}}}
\newunicodechar{⇒}{\threecharwidth{$\implies$}}
\newunicodechar{≡}{\onecharwidth{$\equiv$}}
\newunicodechar{∖}{\onecharwidth{$\setminus$}}

\newunicodechar{ℤ}{\onecharwidth{$\mathbb Z$}}
\newunicodechar{ℕ}{\onecharwidth{$\mathbb N$}}
\newunicodechar{𝕄}{\onecharwidth{\ensuremath{\mathbb M}}}
\newunicodechar{β}{\onecharwidth{$\beta$}}
\newunicodechar{ϕ}{\onecharwidth{$\phi$}}

\newunicodechar{ⵌ}{\onecharwidth{${{}^\#}$}}
\newunicodechar{⊕}{\onecharwidth{$\oplus$}}
\newunicodechar{ }{\,}
\newunicodechar{ }{\,}

\newunicodechar{⊤}{\onecharwidth{$\top$}}
\newunicodechar{⊥}{\onecharwidth{$\bot$}}
\newunicodechar{λ}{\onecharwidth{$\lambda$}}
\newunicodechar{∨}{\onecharwidth{$\vee$}}
\newunicodechar{∧}{\onecharwidth{$\wedge$}}
\newunicodechar{≠}{\onecharwidth{$\neq$}}
\newunicodechar{∃}{\onecharwidth{$\exists$}}
\newunicodechar{∀}{\onecharwidth{$\forall$}}
\newunicodechar{×}{\onecharwidth{$\times$}}
\newunicodechar{≤}{\onecharwidth{$\leq$}}
\newunicodechar{≥}{\onecharwidth{$\geq$}}
\newunicodechar{⇔}{\threecharwidth{$\iff$}}
\newunicodechar{∇}{\threecharwidth{$\nabla$}}

\newunicodechar{⦇}{\onecharwidth{$\llparenthesis$}}
\newunicodechar{⦈}{\onecharwidth{$\rrparenthesis$}}

\newunicodechar{₀}{\onecharwidth{${}_0$}}
\newunicodechar{₁}{\onecharwidth{${}_1$}}
\newunicodechar{₂}{\onecharwidth{${}_2$}}
\newunicodechar{₃}{\onecharwidth{${}_3$}}
\newunicodechar{₄}{\onecharwidth{${}_4$}}
\newunicodechar{₅}{\onecharwidth{${}_5$}}
\newunicodechar{₆}{\onecharwidth{${}_6$}}
\newunicodechar{₇}{\onecharwidth{${}_7$}}
\newunicodechar{₈}{\onecharwidth{${}_8$}}
\newunicodechar{₉}{\onecharwidth{${}_9$}}
\newunicodechar{ⁿ}{\onecharwidth{${}^n$}}
\newunicodechar{ₙ}{\onecharwidth{${}_n$}}
\newunicodechar{ₘ}{\onecharwidth{${}_m$}}

\newcommand*\circled[1]{\tikz[baseline=-1mm]{
    \node[shape=circle,fill,draw,inner sep=1pt] (char) {\tiny\rm\bf\color{white}#1};}}
\newunicodechar{①}{\circled{1}}
\newunicodechar{②}{\circled{2}}
\newunicodechar{③}{\circled{3}}
\newunicodechar{④}{\circled{4}}
\newunicodechar{⑤}{\circled{5}}
\newunicodechar{⑥}{\circled{6}}
\newunicodechar{⑦}{\circled{7}}
\newunicodechar{⑧}{\circled{8}}
\newunicodechar{⑨}{\circled{9}}
\newunicodechar{⑩}{\circled{10}}
\newunicodechar{⑪}{\circled{11}}
\newunicodechar{⑫}{\circled{12}}
\newunicodechar{⑬}{\circled{13}}
\newunicodechar{⑭}{\circled{14}}
\newunicodechar{⑮}{\circled{15}}
\newunicodechar{⑯}{\circled{16}}
\newunicodechar{⑰}{\circled{17}}
\newunicodechar{ⓑ}{\circled{b}}
\newunicodechar{٪}{\%}
\newunicodechar{…}{.\!.\!.}
\newunicodechar{­}{\-}

\makeatletter
\AtBeginEnvironment{minted}{\dontdofcolorbox}
\def\dontdofcolorbox{\renewcommand\fcolorbox[4][]{##4}}
\xpatchcmd{\inputminted}{\minted@fvset}{\minted@fvset\dontdofcolorbox}{}{}
\xpatchcmd{\mintinline}{\minted@fvset}{\minted@fvset\dontdofcolorbox}{}{} 
\makeatother

\usepackage{glossaries}
\usepackage{csquotes}
\makeglossaries

\newglossaryentry{verasco}
{
    name=Verasco,
    description={}
}
 
\newglossaryentry{metastar}
{
    name=Meta-F$^\star$,
    description={}
}

\newglossaryentry{fstar}
{
    name=F$^\star$,
    description={}
}

\newglossaryentry{lowstar}
{
    name=Low$^\star$,
    description={}
}

\newglossaryentry{csharpminor}
{
    name=C\#minor,
    description={}
}

\newglossaryentry{wp}
{
        name=weakest-precondition,
        description={}
}

\newglossaryentry{bind}
{
    name=bind,
    description={}
}
\newglossaryentry{pure}
{
    name=pure,
    description={}
}

\LetLtxMacro{\cminted}{\minted}

\xpretocmd{\cminted}{\RecustomVerbatimEnvironment{Verbatim}{BVerbatim}{}}{}{}

\newmintinline[fcode]{./latex/fstar.py -x}{escapeinside=!!,mathescape=true}%
\newmintinline[fcodemm]{./latex/fstar.py -x}{escapeinside=!!,mathescape=true}%
\newminted[fstarcode]{./latex/fstar.py -x}{autogobble,escapeinside=!!,mathescape=true}%
\newmintedfile[inputfstar]{./latex/fstar.py -x}{autogobble,escapeinside=!!,mathescape=true}%

\LetLtxMacro{\cfstarcode}{\fstarcode}

\xpretocmd{\cfstarcode}{\RecustomVerbatimEnvironment{Verbatim}{BVerbatim}{}}{}{}

\newcommand{\shorteq}{\resizebox{4pt}{\height}{=}}
\newcommand{\shortgt}{\resizebox{4pt}{\height}{\textgreater}}
\newcommand{\bindSymbol}{%
   \textrm{\shortgt\kern-2pt\shortgt\kern-1pt\shorteq}%
}

\def\fEQlabel{}%
\newif\ifflagFEQLabel%
\makeatletter
\newcommand{\customlabel}[2]{%
   \protected@write \@auxout {}{\string \newlabel {#1}{{#2}{\thepage}{#2}{#1}{}} }%
   \hypertarget{#1}{}%
}
\makeatother
\LetLtxMacro{\fequation}{\cfstarcode}%
\let\endfequation\endcfstarcode%
\xpretocmd{\fequation}{%
\noindent%
  \list{}{\rightmargin=0pt\leftmargin=0pt}\item\relax%
  \LetLtxMacro{\savelabel}{\label}%
  \vspace{5.5pt}%
  \flagFEQLabelfalse%
  \ifdefempty{\fEQlabel}{}{\flagFEQLabeltrue}%
  \ifflagFEQLabel%
  \stepcounter{equation}%
  \customlabel{\fEQlabel}{\theequation}%
  \minipage{1\linewidth}%
  \fi%
  \centering%
}{}{}%
\xapptocmd{\endfequation}{%
  \ifflagFEQLabel%
  \endminipage%
  \kern-0.6cm\minipage{0.04\linewidth}%
  \flushright%
  \eqref{\fEQlabel}%
  \endminipage%
  \fi%
  \vspace{5.5pt}%
  \endlist%
  \def\fEQlabel{}%
  \global\def\fEQlabel{}%
}{}{}%

\def\AbIntImp{\textrm{IMP}}

\title{Verified Functional Programming \\of an Abstract Interpreter}

\usepackage{cleveref}
\newcommand{\orcidlink}[1]{\href{https://orcid.org/#1}{\includegraphics[height=0.6em]{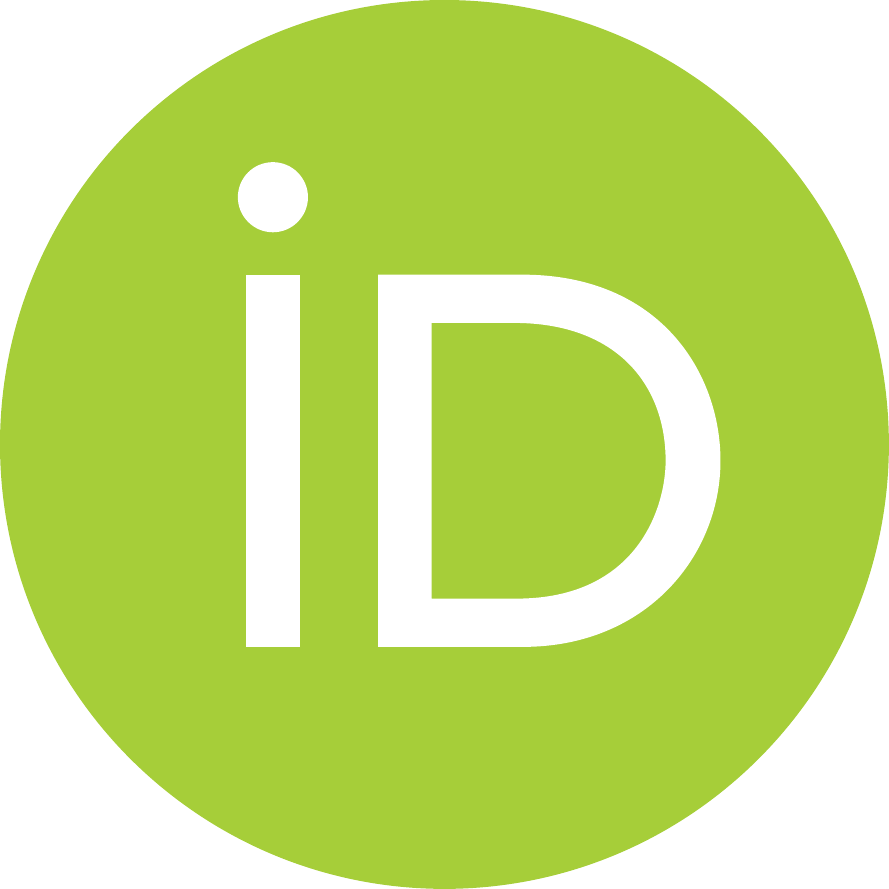}}}

\begin{document}
\author{%
Lucas Franceschino\inst{1}
\,\orcidID{0000-0002-5683-0199}
\and %
David Pichardie\inst{2}
\,\orcidID{0000-0002-2504-1760}
\and %
Jean-Pierre Talpin\inst{1}
\,\orcidID{0000-0002-0556-4265}
}
\institute{%
${}^1$~INRIA Rennes, France %
~~~${}^2$~ENS Rennes, France%
}

\maketitle{}
\begin{abstract}
Abstract interpreters are complex pieces of software: even if the
abstract interpretation theory and companion algorithms are well
understood, their implementations are subject to bugs, that might
question the soundness of their computations.

While some formally verified abstract interpreters have been written
in the past, writing and understanding them requires expertise in the
use of proof assistants, and requires a non-trivial amount of
interactive proofs.

This paper presents a formally verified abstract interpreter fully
programmed and proved correct in the F* verified programming
environment. Thanks to F* refinement types and SMT prover capabilities
we demonstrate a substantial saving in proof effort compared to
previous works based on interactive proof assistants.
Almost all the code of our implementation, proofs included, written in
a functional style, are presented directly in the paper.

\end{abstract}

\begin{center}
 \includegraphics[scale=0.64]{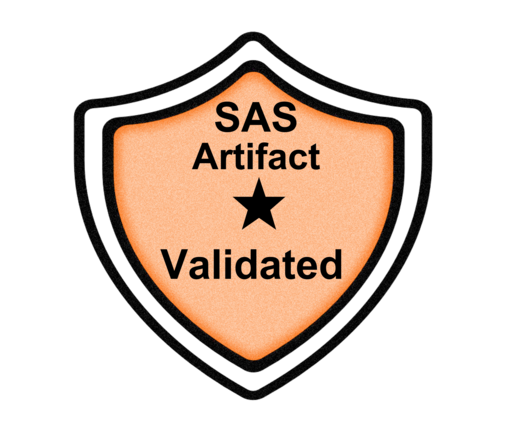} %
 \includegraphics[scale=0.64]{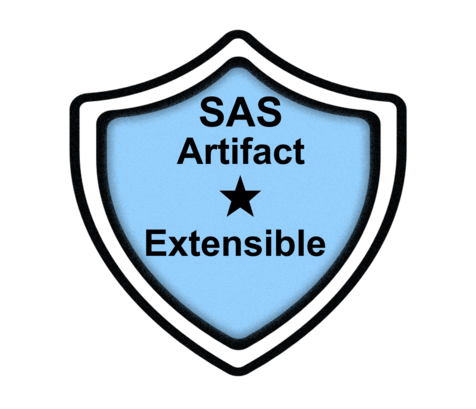} %
 \includegraphics[scale=0.64]{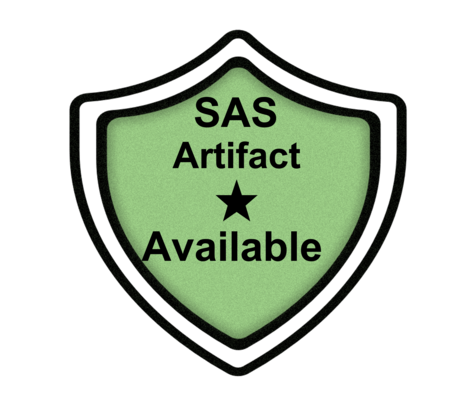}%
\end{center}
\section{Introduction}

Abstract interpretation is a theory of sound approximation.
However, most of available abstract interpreters do not formally
establish a relation between their algorithmic theory and
implementations.
Several abstract interpreters have been proven correct. The most
notable one is Verasco~\cite{POPL15:Jourdan:al}, a static analyser of C
programs that has been entirely written, specified and proved in the
proof assistant Coq.
However, understanding the implementation and proof of Verasco
requires an expertise with Coq and proof assistants.

Proofs in Coq are achieved thanks to extensive use of proof scripts,
that are very difficult for non expert to read.
By contrast with a handwritten proof, a Coq proof can be very verbose,
and often does not convey a good intuition for the idea behind a
proof.
Thus, writing and proving sound a static analyzer is a complex and
time-consuming task: for example, Verasco requires about 17k
lines~\cite{POPL15:Jourdan:al} of manual Coq proofs.
Such an effort, however, yields the strongest guarantees and provides
complete trust in the static analyser.

This paper showcases the implementation of a sound static
analyser using the general-purpose functional programming language
\gls{fstar}.
Equipped with dependent types and built-in SMT solver facilities,
\gls{fstar} provides both an OCaml-like experience and proof assistant
capacities. It recently shined with the Project
Everest~\cite{project-everest}, which delivered a series of formally
verified, high-performance, cryptographic libraries:
HACL*~\cite{haclstar}, ValeCrypt~\cite{bond2017vale} and
EverCrypt~\cite{protzenko2020evercrypt}; that are for instance used and
deployed in Mozilla Firefox.
While \gls{fstar} can always resort to proof scripts similar to Coq's
ones, most proof obligations in \gls{fstar} are automatically
discharged by the SMT solver Z3~\cite{z3}.

We present an abstract interpreter equipped with the numerical
abstract domain of intervals, forward and backward analyses of
expressions, widening, and syntax-directed loop iteration.
This paper makes the following contributions.

\begin{itemize}
\item It demonstrates the ease of use of \gls{fstar} for verified
  static analysis: we implement a verified abstract interpreter, and
  show about 95\% of its 527 lines of code (proof included) directly
  in the paper.
\item
  As far as we know, it is the first time SMT techniques are used for
  verifying an abstract interpreter.
\item
  We gain an order of magnitude in the number of proof lines in
  comparison with similar works implemented in Coq.
  %
  %

\end{itemize}

\paragraph{Related work} Efforts in verified abstract interpretation are
numerous~\cite{Pich:these,ITP10:Cachera:Pichardie,SAS13:Blazy:al,Nipkow-ITP12},
and go up to Verasco~\cite{POPL15:Jourdan:al}, a modular, real-world
abstract interpreter verified in Coq.
Blazy et al.~\cite{SAS13:Blazy:al} and Verasco follow closely the
modular design of Astrée~\cite{astree}; we exhibit a similar modularity
on a smaller scale.
However, such analysers require a non-trivial amount of mechanized
proofs: in constrast, this paper shows that implementing a formally
verified abstract interpreter with very little manual proofs is
possible.
So far, verified abstract interpreters have been focused on
concretization-based formalizations.
The work of Darais et al.~\cite{darais-oopsla15} is the only one to
really consider the use of Galois connections.
They provide a minimalist abstract inteperter for imperative language
but this interpreter seems very limited compared to ours.
They use the Agda proof assistant which is comparable to Coq in terms
of proof verbosity.

\paragraph{Overview} Section~\ref{abint:section:lang-def} defines $\AbIntImp{}$, the language our
abstract interpreter deals with, to which is given an operational
semantics in Section~\ref{abint:section:operational-sematics}.
Then Section~\ref{abint:section:abstract-domains} formalizes lattices and abstract
domains, while Section~\ref{abint:section:intervals} instantiates them with the
abstract domain of intervals.
Section~\ref{abint:section:specialized-adoms} derives more specific abstract domains,
for numeric expressions and for memories.
The latter is instantiated by Section~\ref{abint:section:weakly-rel-amem}, that implements an
abstract weakly-relational memory.
Finally, Section~\ref{abint:section:stmt-ab-int} presents the abstract interpretation
of $\AbIntImp{}$ statements.

\paragraph{}
The \gls{fstar} development is available on
GitHub\footnote{\url{https://github.com/W95Psp/verified-abstract-interpreter}}
or as supplementary material~\cite{supp-mat}.
The resulting analyser is available online as a web application at
\url{https://w95psp.github.io/verified-abstract-interpreter}.

\section{$\AbIntImp{}$: a Small Imperative Language}
\label{abint:section:lang-def}
To present our abstract intrepreter, we first show the language on
which it operates: $\AbIntImp{}$. It is a simple imperative language,
equipped with memories represented as functions from variable names
\fcode{varname} to signed integers, \fcode{int_m}.
This presentation lets the reader unfamiliar with \gls{fstar} get
used to its syntax: $\AbIntImp{}$'s \gls{fstar} definition looks like
OCaml; the main difference is the explicit type signatures for
constructors in algebraic data types.
$\AbIntImp{}$ has numeric expressions, encoded by the type
\fcode{expr}, and statements \fcode{stmt}.
Booleans are represented numerically: $0$ represents
\fcode{false}, and any other value stands for true.
The enumeration \fcode{binop} equips $\AbIntImp{}$ with
various binary operations. The constructor \fcode{Unknown} encodes an
arbitrary number.
Statements in $\AbIntImp{}$ are the assignment, the
non-deterministic choice, the sequence and the loop.
\begin{fstarcode}
type varname = | VA | VB | VC | VD   type mem 'a = varname -> 'a
type binop = | Plus | Minus | Mult | Eq | Lt | And | Or
type expr = | Const:  int_m -> expr   | Var: varname -> expr
            | BinOp: binop -> expr -> expr -> expr | Unknown
type stmt = | Assign: varname -> expr -> stmt | Assume: expr -> stmt
            | Seq:       stmt -> stmt -> stmt | Loop:   stmt -> stmt
            | Choice: stmt -> stmt -> stmt
\end{fstarcode}
The type \fcode{int_m} is a \emph{refinement} of the built-in
\gls{fstar} type \fcode{int}: while every integer lives in the
type \fcode{int}, only those that respect certain bounds live in
\fcode{int_m}.
Numerical operations (\fcode{+}, \fcode{-} and \fcode{*}) on
machine integers wrap on overflow, i.e. adding one to the maximal
machine integer results in the minimum machine integer.
We do not give the detail of their implementation.
\section{Operational Semantics}
\label{abint:section:operational-sematics}
This section defines an operational semantics for $\AbIntImp{}$. It is
also a good way of introducing more \gls{fstar} features.

We choose to formulate our semantics in terms of sets.
Sets are encoded as maps from values to propositions~\fcode{prop}.
Those are logical statements and shouln't be confused with
booleans.
Below, \fcode{⊆} quantifies over every \emph{inhabitant} of a
type: stating whether such a statement is true or false is clearly
not computable.
Arbitrarily complex properties can be expressed as propositions of
type \fcode{prop}.

In the listing below, notice the greek letters: we use them
throughout the paper.
They denote implicit type arguments: for instance, below, $\in$
works for any set \fcode{set 'a}, with any type \fcode{'a}.
\gls{fstar} provides the propositional operators $\wedge$, $\vee$
and \fcode{==}, in addition to boolean
ones~(\fcode{&&},~\fcode{||} and~\fcode{=}).
We use them below to define the union, intersection and
differences of sets.
\begin{fstarcode}
type set 'a = 'a -> prop             let (∈) (x: 'a) (s: set 'a) = s x
let (∩) s0 s1 = fun x -> x ∈ s0 /\ x ∈ s1  let (∖) s0 s1 = fun v -> s0 v /\ ~(s1 v)
let (∪) (s0 s1: set 'a): set 'a = fun x -> x ∈ s0 \/ x ∈ s1
let (⊆) (s0 s1: set 'a): prop = forall (x: 'a). x ∈ s0 ==> x ∈ s1
let set_inverse (s: set int_m): set int_m = fun (i: int_m) -> s (-i)
\end{fstarcode}
To be able to work conveniently with binary operations on integers
in our semantics, we define \fcode{lift_binop}, that lifts them as
set operations.
For example, the set \fcode{lift_binop (+) a b} (\fcode{a} and
\fcode{b} being two sets of integers) corresponds to
$\left\{
va + vb \mid va\in \textrm{\fcode{a}} \wedge vb \in \textrm{\fcode{b}}
\right\}$.
\begin{fstarcode}
let lift_binop (op: 'a -> 'a -> 'a) (a b: set 'a): set 'a
  = fun r -> exists (va:'a). exists (vb:'a). va ∈ a /\ vb ∈ b /\ r == op va vb
unfold let lift op = lift_binop (concrete_binop op)
\end{fstarcode}
The binary operations we consider are enumerated by
\fcode{binop}. The function \fcode{concrete_binop} associates these syntactic operations to integer operations.
For convenience, \fcode{lift} maps a \fcode{binop} to a set
operation, using \fcode{lift_binop}. This function is inlined by
\gls{fstar} directly when used because of the keyword
\fcode{unfold}; intuitively \fcode{lift} behaves as a macro.
\begin{fstarcode}
unfold let concrete_binop (op: binop): int_m -> int_m -> int_m
  = match op with | Plus -> n_add | Lt -> lt_m | ... | Or -> ori_m
\end{fstarcode}
The operational semantics for expressions is given as a map from
memories and expressions to sets of integers.
Notice the use of both the syntax \fcode{val} and \fcode{let} for
the function \fcode{osem_expr}.
The \fcode{val} syntax gives \fcode{osem_expr} the type
\fcode{mem->expr->set int_m}, while the \fcode{let} declaration
gives its definition.
The semantics itself is uncomplicated: \fcode{Unknown} returns
the set of every \fcode{int_m}, a constant or a \fcode{Var}
returns a singleton set. For binary operations, we lift them as
set operations, and make use of recursion.
\begin{fstarcode}
val osem_expr: mem -> expr -> set int_m
let rec osem_expr m e = fun (i: int_m) 
  -> match e with | Const x -> i==x | Var v -> i==m v | Unknown -> True
  | BinOp op x y -> lift op (osem_expr m x) (osem_expr m y) i
\end{fstarcode}
The operational semantics for statements maps a statement and an
initial memory to a set of admissible final memories.
Given a statement \fcode{s}, an initial memory \fcode{m_i} and a
final one \fcode{m_f}, \fcode{osem_stmt s m_i m_f}
(defined below) is a proposition stating whether the
transition is possible.
\begin{fstarcode}
val osem_stmt (s: stmt): mem -> set mem
let rec osem_stmt (s: stmt) (m_i m_f: mem)
  = match s with
  | Assign v e -> ∀w. if v = w then m_f v ∈ osem_expr m_i e
                              else m_f w == m_i w
  | Seq a b  -> exists (m1: mem). m1 ∈ osem_stmt a m_i /\ m_f ∈ osem_stmt b m1
  | Choice a b -> m_f ∈ (osem_stmt a m_i ∪ osem_stmt b m_i)
  | Assume e -> m_i == m_f /\ (exists (x: int_m). x <> 0 /\ x ∈ osem_expr m_i e)
  | Loop a -> closure (osem_stmt a) m_i m_f
\end{fstarcode}
The simplest operation is the assignment of a variable \fcode{v}
to an expression \fcode{e}: the transition is allowed if every
variable but \fcode{v} in \fcode{m_i} and \fcode{m_f} is equal and
the final value of \fcode{v} matches with the semantics of
\fcode{e}.
Assuming that an expression is true amounts to require the initial
memory to be such that at least a non-zero integer (that is, the
encoding of \fcode{true}) belongs to \fcode{osem_expr m_i e}.
The statement \fcode{Seq a b} starting from the initial memory
\fcode{m_i} admits \fcode{m_f} as a final memory when there exists
\begin{enumi} \item a transition from \fcode{m_i} to an
intermediate memory \fcode{m1} with statement \fcode{a} and \item
a transition from \fcode{m1} to \fcode{m_f} with statement
\fcode{b}.  \end{enumi}
The operational semantics for a loop is defined as the reflexive
transitive closure of the semantics of its body.
The \fcode{closure} function computes such a closure, and is
provided by \gls{fstar}'s standard library.
\section{Abstract Domains}
\label{abint:section:abstract-domains}
Our abstract interpreter is parametrized over relational domains.
We instantiate it later with a weakly-relational~\cite{astree}
memory.
This section defines lattices and abstract domains.
Such structures are a natural fit for typeclasses~\cite{metafstar},
which allow for ad hoc polymorphism.
In our case, it means that we can have one abstraction for
lattices for instance, and then instantiate this abstraction with
implementations for, say, sets of integers, then intervals, etc.
Typeclasses can be seen as record types with dedicated dependency
inference.
Below, we define the typeclass \fcode{lattice}: defining an
instance for a given type equips this type with a lattice
srtucture.
\paragraph{Refinement types}
Below, the syntax \fcode{x:'t{p x}} denotes the type whose
inhabitants both belong to \fcode{'t} and satisfy the predicate
\fcode{p}.
For example, the inhabitant of the type
\fcode{bot:nat{forall(n:nat).  bot <= n}} is \fcode{0}: it is the
(only) smallest natural number.
To typecheck \fcode{x:'t}, \gls{fstar} collects the \textit{proof
obligations} implied by "\fcode{x} has the type \fcode{'t}", and
tries to discharge them with the help of the SMT solver.
If the SMT solver is able to deal with the proof obligations, then
\fcode{x:'t} typechecks.
In the case of "\fcode{0} is of type
\fcode{bot:nat{forall(n:nat).  bot <= n}}", the proof obligation
is \fcode{forall(n:nat). 0 <= n}.

Below, most of the types of the fields from the record type
 \fcode{lattice} are refined.
Typechecking \fcode{i} against the type \fcode{lattice 'a} yields
a proof obligation asking (among other things) for \fcode{i.join}
to go up in the lattice and for \fcode{bottom} to be a lower
bound.
Thus, if "\fcode{i} has type \fcode{lattice 'a}" typechecks, it
means there exists a proof that the properties written as
refinements in \fcode{lattice}'s definition hold on \fcode{i}.
We found convenient to let \fcode{bottom} represent unreachable
states.
Note \fcode{lattice} is under-specified, i.e. it doesn't require
\fcode{join} to be provably a least upper bound, since such a
property plays no role in our proof of soundness. This choice
follows Blazy and al.~\cite{SAS13:Blazy:al}.
\begin{fstarcode}
class lattice 'a = { corder: order 'a
  ; join: x:'a -> y:'a -> r:'a {corder x r /\ corder y r}
  ; meet: x:'a -> y:'a -> r:'a {corder r x /\ corder r y}
  ; bottom:  bot:'a {∀x. corder bot x}; top:  top:'a {∀x. corder x top}}
\end{fstarcode}
For our purpose, we need to define what an abstract domain is. In
our setting, we consider concrete domains with powerset structure.
The typeclass \fcode{adom} encodes them: it is parametrized by a
type \fcode{'a} of abstract values.
For instance, consider \fcode{itv} the type for intervals:
\fcode{adom itv} would be the type inhabited by correct abstract
domains for intervals.

Implementing an abstract domain amounts to implementing the
following fields: \begin{enumi}
\item \fcode{c}, that represents the type to which abstract values
\fcode{'a} concretizes;
\item \fcode{adom_lat}, a lattice for \fcode{'a};
\item \fcode{widen}, a widening operator;
\item \fcode{gamma}, a monotonic concretization function from \fcode{'a}
to \fcode{set c};
\item \fcode{order_measure}, a measure ensuring the abstract
domain doesn't admit infinite increasing chains, so that
termination is provable for fixpoint iterations;
\item \fcode{meet_law}, that requires \fcode{meet} to be a correct
approximation of set intersection;
\item \fcode{top_law} and \fcode{bot_law}, that ensure the lattice's
bottom concretization matches with the empty set, and similarly
for \fcode{top}.
\end{enumi}
\begin{fstarcode}
class adom 'a = { c: Type; adom_lat: lattice 'a
  ; gamma: (gamma: ('a -> set c) {forall (x y: 'a). corder x y ==> (gamma x ⊆ gamma y)})
  ; widen: x:'a -> y:'a -> r:'a {corder x r /\ corder y r}
  ; order_measure: measure adom_lat.corder
  ; meet_law: x:'a -> y:'a -> Lemma ((gamma x ∩ gamma y) ⊆ gamma (meet x y))
  ; bot_law: unit -> Lemma (forall (x:c). ~(x ∈ gamma bottom))
  ; top_law: unit -> Lemma (forall (x:c). x ∈ gamma top)}
\end{fstarcode}
Notice the refinement types: we require for instance the monotony
of \fcode{gamma}.
Every single instance for \fcode{adom} will be checked against
these specifications. No instance of \fcode{adom} where
\fcode{gamma} is not monotonic can exist.
With a proposition \fcode{p}, the \fcode{Lemma p} syntax signals a
function whose outcome is computationally irrelevant, since it
simply produces \fcode{()}, the inhabitant of the type
\fcode{unit}.
However, it does not produces an arbitrary \fcode{unit}: it
produces an inhabitant of \fcode{_:unit {p}}, that is, the type
\fcode{unit} refined with the goal \fcode{p} of the lemma itself.

For praticity, we define some infix operators for \fcode{adom_lat}
functions.
The syntax \fcode{{|...|}} lets one formulate typeclass
constraints: for example, \fcode{(⊑)} below ask \gls{fstar} to
resolve an instance of the typeclass \fcode{adom} for the type
\fcode{'a}, and name it \fcode{l}.
Below, \fcode{(⊓)} instantiates the lemma \fcode{meet_law}
explicitly: \fcode{meet_law x y} is a unit value that carries a
proof in the type system.
\begin{fstarcode}
let (⊑) {|l:adom 'a|} = l.adom_lat.corder
let (⊔) {|l:adom 'a|} (x y:'a): r:'a { corder x r /\ corder y r 
                                 /\ (gamma x ∪ gamma y) ⊆ gamma r } = join x y
let (⊓) {|l:adom 'a|} (x y:'a): r:'a { corder r x /\ corder r y
                                 /\ (gamma x ∩ gamma y) ⊆ gamma r }
  = let _ = meet_law x y in meet x y
\end{fstarcode}
Lemmas are functions that produce refined \fcode{unit} values
carrying proofs.
Below, given an abstract domain \fcode{i}, and two abstract values
\fcode{x} and \fcode{y}, \fcode{join_lemma i x y} is a proof
concerning \fcode{i}, \fcode{x} and \fcode{y}.
Such an instantiation can be manual (i.e. below, \fcode{i.top_law
()} in \fcode{top_lemma}), or automatic.
The automatic instantiation of a lemma is decided by the
SMT solver.
Below, we make use of the \fcode{SMTPat} syntax, that allows us to
give the SMT solver a list of patterns.
Whenever the SMT solver matches a pattern from the list, it
instantiates the lemma in stake.
The lemma \fcode{join_lemma} below states that the union of the
concretization of two abstract values \fcode{x} and \fcode{y} is
below the concretization of the abstract join of \fcode{x} and
\fcode{y}.
This is true because of \fcode{gamma}'s monotony: we help a bit
the SMT solver by giving a hint with \fcode{assert}.
This lemma is instantiated each time a proof goal contains
\fcode{x ⊑ y}.

Because of a technical limitation, we cannot write SMT patterns
directly in the \fcode{meet_law}, \fcode{bot_law} and
\fcode{top_law} fields of the class \fcode{adom}: thus, below we
reformulate them.
\begin{fstarcode}
let top_lemma (i: adom 'a)      (let bot_lemma, meet_lemma = ...)
  : Lemma (forall (x: i.c). x ∈ i.gamma i.adom_lat.top)
          [SMTPat (i.gamma i.adom_lat.top)] = i.top_law ()
let join_lemma (i: adom 'a) (x y: 'a)
  : Lemma ((i.gamma x ∪ i.gamma y) ⊆ i.gamma (i.adom_lat.join x y))
          [SMTPat (i.adom_lat.join x y)]
  = let r = i.adom_lat.join x y in assert (gamma x ⊆ gamma r /\ gamma y ⊆ gamma r)
\end{fstarcode}
\section{An Example of Abstract Domain: Intervals}
\label{abint:section:intervals}
Until now, the \gls{fstar} code we presented was mostly
specificational.
This section presents the abstract domain of intervals, and thus
shows how proof obligations are dealt with in \gls{fstar}.
Below, the type \fcode{itv'} is a dependent tuple: the refinement
type on its right-hand side component \fcode{up} depends on
\fcode{low}.
If a pair \fcode{(|x,y|)} is of type \fcode{itv'}, we have a proof
that \fcode{x <= y}.
\begin{fstarcode}
type itv' = low:int_m & up:int_m {low <= up}    type itv  =  withbot itv'
\end{fstarcode}
The machine integers being finite, \fcode{itv'} naturally has a
top element.
However, \fcode{itv'} cannot represent the empty set of integers,
whence \fcode{itv}, that adds an explicit bottom element using
\fcode{withbot}.
The syntax \fcode{Val?} returns true when a value is not
\fcode{Bot}.
For convenience, \fcode{mk} makes an interval out of two numbers,
and \fcode{itv_card} computes the cardinality of an interval. We
use it later to define a measure for intervals.
\fcode{inbounds x} holds when \fcode{x:int} fits machine integer
bounds.
\begin{fstarcode}
type withbot (a: Type) = | Val: v:a -> withbot a | Bot
let mk (x y: int): itv = if inbounds x && inbounds y && x <= y
                       then Val (|x,y|) else Bot
let itv_card (i:itv):nat = match  i  with | Bot -> 0 | Val i -> dsnd i - dfst i + 1
\end{fstarcode}
Below, \fcode{lat_itv} is an instance of the typeclass
\fcode{lattice} for intervals: intervals are ordered by inclusion,
the \fcode{meet} and \fcode{join} operations consist in unwrapping
\fcode{withbot}, then playing with bounds.
\fcode{lat_itv} is of type \fcode{lattice itv}: it means for
instance that we have the proof that the join and meet operators
respect the order \fcode{lat_itv.corder}, as stated in the
definition of \fcode{lattice}.
Note that here, not a single line of proof is required:
\gls{fstar} transparently builds up proof obligations, and asks
the SMT to discharge them, that does so automatically.
\begin{fstarcode}
instance lat_itv: lattice itv =
  { corder = withbot_ord #itv' (fun (|a,b|) (|c,d|) -> a>=c && b<=d)
  ; join = (fun (i j: itv) -> match i, j with
       | Bot, k | k, Bot -> k
       | Val (|a,b|), Val (|c,d|) -> Val (|min a c, max b d|))
  ; meet = (fun (x y: itv) -> match x, y with
       | Val (|a,b|), Val (|c,d|) -> mk (max a c) (min b d)
       | _ -> Bot); bottom = Bot; top = mk min_int_m max_int_m }
\end{fstarcode}
Such automation is possible even with more complicated
definitions: for instance, below we define the classical widening
with thresholds.
Without a single line of proof, \fcode{widen} is shown as
respecting the order \fcode{corder}.
\begin{fstarcode}
let thresholds: list int_m  =  [min_int_m;-64;-32;-16;-8;-4;4;8;16;32...]
let widen_bound_r (b: int_m): (r:int_m {r>b \/ b=max_int_m}) =
  if b=max_int_m then b else find' (fun (u:int_m) -> u>b) thresholds
let widen_bound_l (b: int_m): (r:int_m {r<b \/ b=min_int_m}) =
  if b=min_int_m then b else find' (fun (u:int_m) -> u<b) (rev thresholds)
let widen (i j: itv): r:itv {corder i r /\ corder j r}
  = match i, j with | Bot, x | x, Bot -> x
  | Val(|a,b|),Val(|c,d|) -> Val (| (if a <= c then  a  else  widen_bound_l  c)
                              , (if b >= d then  b  else  widen_bound_r  d)|)
\end{fstarcode}
Similarly, turning \fcode{itv} into an abstract domain requires
no proof effort.
Below \fcode{itv_adom} explains that intervals concretize to
machine integers (\fcode{c = int_m}), how it does so (with
\fcode{gamma = itv_gamma}), and which lattice is associated with
the abstract domain (\fcode{adom_lat = lat_itv}).
As explained previously, the proof of a proposition $p$ in F* can
be encoded as an inhabitant of a refinement of \fcode{unit},
whence the "empty" lambdas: we let the SMT solver figure out the
proof on its own.
\begin{fstarcode}
let itv_gamma: itv -> set int_m =  withbot_gamma  (fun (i:itv') x -> dfst i <= x /\ x <= dsnd i)
instance itv_adom: adom itv = { c = int_m   ; adom_lat = lat_itv; gamma = itv_gamma
    ; meet_law = (fun _ _->()); bot_law = (fun _->());  top_law = (fun _->())
    ; widen = widen     ; order_measure={f=itv_card;max=size_int_m}}
\end{fstarcode}
\subsection{Forward Binary Operations on Intervals}
\label{abint:sub:forward-op-itv}
Most of the binary operations on intervals can be written and
shown correct without any proof. Our operators handle machine
integer overflowing: for instance, \fcode{add_overflows} returns a
boolean indicating whether the addition of two integers overflows,
solely by performing machine integer operations.
The refinement of \fcode{add_overflows} states that the returned
boolean \fcode{r} should be true if and only if the addition in
\fcode{int} differs from the one in \fcode{int_m}.
The correctness of \fcode{itv_add} is specified as a refinement:
the set of the additions between the concretized values from the
input intervals is to be included in the concretization of the
abstract addition.
Its implementation is very simple, and its correctness proved
automatically.
\begin{fstarcode}
let add_overflows (a b: int_m)
  : (r: bool {r <==> int_arith.n_add a b <> int_m_arith.n_add a b})
  = ((b<0) = (a<0)) && abs a > max_int_m - abs b
let itv_add (x y: itv): (r: itv {(gamma x + gamma y) ⊆ gamma r})
  = match x, y with | Val (|a, b|), Val (|c, d|)
                    -> if add_overflows a c || add_overflows b d
                      then top else Val (|a + c, b + d|) | _->Bot
\end{fstarcode}
However the SMT solver sometimes misses some necessary lemmas.
In such cases, we can either guide the SMT solver by
discriminating cases and inserting hints, or go fully manual with
a tactic system à la Coq.
Below, the \fcode{assert} uses tactics: everything within the
parenthesis following the \fcode{by} keyword is a computation that
manipulates proof goals.
Our aim is to prove that subtracting two numerical sets $a$ and
$b$ is equivalent to adding $a$ with the inverse of $b$.

\sloppy Unfortunately, due to the nature of \fcode{lift_binop},
this yields existential quantifications which are difficult for
the SMT solver to deal with. After normalizing our goal (with
\fcode{compute ()}), and dealing with quantifiers and implications
(\fcode{forall_intro}, \fcode{implies_intro} and
\fcode{elim_exists}), we are left with
\fcode{∃y. b (-y) /\ r=x+y}
knowing \fcode{b z /\ r=x-z} given some \fcode{z} as an
hypothesis.
Eliminating \fcode{∃y} with \fcode{-z} is enough to complete the
proof.

We sadly had to prove that (not too complicated) fact by hand.
This however shows the power of \gls{fstar}. Its type system is
very expressive: one can state arbitrarily mathematically hard
propositions (for which automation is hopeless).
In such cases, one can always resort to Coq-like manual proving to
handle hard proofs.
\begin{fstarcode}
let set_inverse (s: set int_m): set int_m = fun (i: int_m) -> s (-i)
let lemma_inv (a b: set int_m)
  : Lemma ((a-b) ⊆ (a+set_inverse b)) [SMTPat (a+set_inverse b)]
  = assert ((a-b) ⊆ (a+set_inverse b)) by (  compute ();
      let _= forall_intro () in let p0 = implies_intro () in
      let witX,p1 = elim_exists (binder_to_term p0) in
      let witY,p1 = elim_exists (binder_to_term p1) in
      let z: int = unquote (binder_to_term witY) in
      witness witX; witness (quote (-z)))
\end{fstarcode}
Notice the SMT pattern: the lemma \fcode{lemma_inv} will be
instantiated each time the SMT deals with an addition involving an
inverse.
Defining the subtraction \fcode{itv_sub} is a breeze: it simply
performs an interval addition and an interval inversion.
Here, no need for a single line of proof for its correctness
(expressed as a refinement).
\begin{fstarcode}
let itv_inv (i: itv): (r: itv {set_inverse (gamma i) ⊆ gamma r})
  = match i with | Val(|lower, upper|) -> Val(|-upper, -lower|) | _ -> i
let itv_sub (x y:itv): (r: itv {(gamma  x - gamma  y) ⊆ gamma  r}) = itv_add x (itv_inv y)
\end{fstarcode}
Proving multiplication sound on intervals requires a lemma which is
not inferred automatically:
\[
\forall x \in [a,b], y\in[b,c].
   \left[
       \mathrm{min}\left(ac,ad,bc,bd\right)
     , \mathrm{max}\left(ac,ad,bc,bd\right)
   \right]
\]
In that case, decomposing that latter lemma into sublemmas
\fcode{lemma_min} and \fcode{lemma_mul} is enough.
Apart from this lemma, \fcode{itv_mul} is free of any proof term.
\begin{fstarcode}
let lemma_min (a b c d: int) (x: int{a <= x /\ x <= b}) (y: int{c <= y /\ y <= d})
  : Lemma (x*y >= a*c \/ x*y >= a*d \/ x*y >= b*c \/ x*y >= b*d) = ()
unfold let in_btw (x: int) (l u: int) = l <= u /\ x >= l /\ x <= u
let lemma_mul (a b c d x y: int)
  : Lemma (requires in_btw x a b /\ in_btw y c d)
    (ensures x*y >= (a*c) `min` (a*d) `min` (b*c) `min` (b*d)
           /\ x*y <= (a*c) `max` (a*d) `max` (b*c) `max` (b*d))
    [SMTPat (x*y); SMTPat (a*c); SMTPat (b * d)]
  = lemma_min a b c d x y; lemma_min (-b) (-a) c d (-x) y
\end{fstarcode}
\begin{fstarcode}
let mul_overflows (a b:int_m):(r:bool{r<>inbounds (int_arith.n_mul a b)})
  = a <> 0 && abs b > max_int_m `div_m` (abs a)
let itv_mul (x y: itv): r:itv {(gamma x * gamma y) ⊆ gamma r}
  = match x, y with
    | Val (|a, b|), Val (|c, d|) ->
        let l = (a*c) `min` (a*d) `min` (b*c) `min` (b*d) in
        let r = (a*c) `max` (a*d) `max` (b*c) `max` (b*d) in
        if mul_overflows a c || mul_overflows a d
        || mul_overflows b c || mul_overflows b d 
        then top else Val (|l, r|)
    | _ -> Bot
\end{fstarcode}
The forward boolean operators for intervals require no proof at
all; here we only give their type signatures.
A function of interest is \fcode{itv_as_bool}: it returns
\fcode{TT} when an interval does not contain 0, \fcode{FF} when
it is the singleton 0, \fcode{Unk} otherwise.
\begin{fstarcode}
let beta (x: int_m): itv = mk x x
let itv_eq (x y:itv): r:itv {(gamma  x `n_eq` gamma  y)  ⊆  gamma  r} =!\shortdots!  let itv_lt=!\shortdots!
let itv_cgamma (i: itv) (x:int_m): r:bool {r <==> itv_gamma i x} =!\shortdots!
let itv_as_bool (x:itv): ubool !\textcolor{gray}{\texttt{ /{}/ with \textbf{type ubool = |Unk|TT|FF}}}!
  = if beta 0=x || Bot?x then FF else if itv_cgamma x 0 then Unk else TT
let itv_andi (x y: itv): (r: itv {(gamma x `n_and` gamma y) ⊆ gamma r})
  = match itv_as_bool x, itv_as_bool y with
  | TT, TT  -> beta 1 | FF, _ | _, FF -> beta 0 | _, _ -> mk 0 1
let itv_ori (x y: itv): (r: itv {(gamma x `n_or` gamma y) ⊆ gamma r}) =!\shortdots!
\end{fstarcode}
\subsection{Backward Operators}
\label{abint:sub:backward-op-itv}
While a forward analysis for expressions is essential, another
powerful analysis can be made thanks to backward operators.
Typically, it aims at extracting information from a test, and at
refining the abstract values involved in this test, so that we
gain in precision on those abstract values.
Given a concrete binary operator \fcode{⊕}, we define
$\overleftarrow{⊕}$ its abstract backward counterpart.
Let three intervals \fcode{xⵌ}, \fcode{yⵌ}, and \fcode{rⵌ}.
$\overleftarrow{⊕}\fcodemm{ xⵌ yⵌ rⵌ}$ tries to find the most
precise intervals \fcode{xⵌⵌ} and
\fcode{yⵌⵌ} supposing
\fcode{gamma xⵌ ⊕ gamma yⵌ ⊆ gamma rⵌ}.
The soundness of $\overleftarrow{⊕}\fcodemm{ xⵌ yⵌ rⵌ}$ can be
formulated as below.
We later generalize this notion of soundness with the type
\fcode{sound_backward_op}, which is indexed by an abstract domain
and a binary operation.
\begin{fequation}
let xⵌⵌ, yⵌⵌ = (!$\overleftarrow{⊕}$!) xⵌ yⵌ rⵌ in
 forall x y. (x ∈ gamma xⵌ /\ y ∈ gamma yⵌ /\ op x y ∈ gamma rⵌ)
    ==> (x ∈ gamma xⵌⵌ /\ y ∈ gamma yⵌⵌ)
\end{fequation}
As the reader will discover in the rest of this section, this
statement of soundness is proved entirely automatically against
each and every backward operator for the interval domain.
For \fcode{op} a concrete operator, \fcode{sound_backward_op itv
op} is inhabited by sound backward operators for \fcode{op} in the
domain of intervals.
If one shows that $\overleftarrow{⊕}$ is of type
\fcode{sound_backward_op itv (+)}, it means exactly that
$\overleftarrow{⊕}$ is a sound backward binary interval operator
for \fcode{(+)}.
The rest of the listing shows how light in proof and OCaml-looking
the backward operations are.
Below, we explain how \fcode{backward_lt} works: it is a bit
complicated because it hides a "\fcode{backward_ge}" operator.
\begin{fstarcode}
let backward_add: sound_backward_op itv n_add = fun x y r -> x ⊓ (r-y), y ⊓ (r-x)
let backward_sub: sound_backward_op itv n_sub = fun x y r -> x ⊓ (r+y), y ⊓ (x-r)
let backward_mul: sound_backward_op itv n_mul = fun x y r -> 
  let h (i j:itv) = (if j=beta 1 then i ⊓ r else i) in h x y, h y x
let backward_eq:  sound_backward_op itv n_eq
  = fun x y r -> match itv_as_bool r with | TT -> x ⊓ y,x ⊓ y | _ -> x,y
let (∖) (x y: itv): (r: itv {(gamma x ∖ gamma y) ⊆ gamma r}) =!\shortdots!
let backward_and: sound_backward_op itv n_and
  = fun x y r -> match itv_as_bool r,itv_as_bool x,itv_as_bool y with
    | FF, TT, _ -> x, y ⊓ beta 0         | FF, _, TT -> x ⊓ beta 0, y
    | TT,  _, _ -> x ∖ beta 0, y ∖ beta 0   | _ -> x, y
let backward_or: sound_backward_op itv n_or
  = fun x y r -> match itv_as_bool r,itv_as_bool x,itv_as_bool y with
    | TT,FF,Unk | TT,FF,FF -> x, y ∖ beta 0 | TT,Unk,FF -> x ∖ beta 0, y
    | FF, _, TT | FF, TT, _ -> x ⊓ beta 0, y ⊓ beta 0 | _ -> x, y
\end{fstarcode}
Let us look at \fcode{backward_lt}.
Knowing whether \fcode{x < y} holds, \fcode{backward_lt} helps us
refining \fcode{x} and \fcode{y} to more precise intervals.
Let \fcode{x} be the interval $[0;\fcodemm{max_int_m}]$, \fcode{y}
be $[5;15]$ and \fcode{r} be $[0;0]$.
Since the singleton $[0;0]$ represents \fcode{false},
\fcode{backward_lt x y r} aims at refining \fcode{x} and \fcode{y}
knowing that \fcode{x < y} doesn't hold, that is, knowing
$\fcodemm{x} \geq \fcodemm{y}$.
In this case, \fcode{backward_lt} finds $\fcodemm{x'} =
[5;\fcodemm{max_int_m}]$ and $\fcodemm{y'} = [5;15]$.
Indeed, when \fcode{r} is $[0;0]$, \fcode{itv_as_bool r} equals to
\fcode{FF}.
Then we rewrite $\neg(\fcodemm{x} < \fcodemm{y})$ either as
$\fcodemm{y} < \fcodemm{x} + 1$ (when \fcode{x} is
\fcode{incrementable}) or as $\fcodemm{y}-1 < \fcodemm{x}$.
In our case, \fcode{x}'s upper bound is \fcode{max_int_m} (the
biggest \fcode{int_m}): x is not incrementable.
Thus we rewrite $\neg([0;\fcodemm{max_int_m}] < [5;15])$ as $[6;16]
< [0;\fcodemm{max_int_m}]$.

Despite of these different case handling, the implementation of
\fcode{backward_lt} required no proof: the SMT solver takes care
of everything automatically.
\begin{fstarcode}
let backward_lt_true (x y: itv)
  = match x, y with | Bot, _ | _, Bot -> x,y
  | Val(|a,b|), Val(|c,d|) -> mk a (min b (d-1)), mk (max (a+1) c) d
let decrementable i=Val?i&&dfst(Val?.v i)>min_int_m  let incr.=...
let backward_lt: sound_backward_op itv n_lt
  = fun x y r -> match itv_as_bool r with | TT -> backward_lt_true x y
  | FF -> if incrementable x // x < y ⇔ y > x+1
         then let ry, rx = backward_lt_true y (itv_add x (beta 1)) in
              itv_sub rx (beta 1), ry
         else if decrementable y // x < y ⇔ y-1 > x
              then let ry, rx = backward_lt_true (itv_sub y (beta 1)) x in
                   rx, itv_add ry (beta 1)
              else x,y | _ -> x, y
\end{fstarcode}
\section{Specialized Abstract Domains}
\label{abint:section:specialized-adoms}
Abstract domains are defined in Section~\ref{abint:section:abstract-domains} as
lattices equipped with a sound concretization operation.
Our abstract interpreter analyses $\AbIntImp{}$ programs: its
expressions are numerical, and $\AbIntImp{}$ is equipped with a memory.
Thus, this section defines two specialized abstract domains: one
for numerical abstractions, and another one for memory
abstractions.
\subsection{Numerical Abstract Domains}
In the section \ref{abint:sub:backward-op-itv}, we explain what a sound
backward operator is in the case of the abstract domain of
intervals.
There, we mention a more generic type
\fcode{sound_backward_op} that states soundness for such operators
in the context of any abstract domain. We present its definition
below:
\begin{fstarcode}
type sound_backward_op (a:Type) {|l:adom a|} (op:l.c->l.c->l.c)
  = backward_op: (a -> a -> a -> (a & a)) {
      forall (xⵌ yⵌ rⵌ: a). let xⵌⵌ, yⵌⵌ = backward_op xⵌ yⵌ rⵌ in
         (forall (x y: l.c). (x ∈ gamma xⵌ /\ y ∈ gamma yⵌ /\ op x y ∈ gamma rⵌ)
                  ==> (x ∈ gamma xⵌⵌ /\ y ∈ gamma yⵌⵌ))}
\end{fstarcode}
We define the specialized typeclass \fcode{num_adom} for abstract
domains that concretize to machine integers.
A type that implements an instance of \fcode{num_adom} should also
have an instance of \fcode{adom}, with \fcode{int_m} as concrete
type.
Whence the fields \fcode{na_adom}, and \fcode{adom_num}.
Moreover, we require a computable concretization function
\fcode{cgamma}, that is, a function that maps abstract values to
computable sets of machine integers: \fcode{int_m -> bool}.
The \fcode{beta} operator lifts a concrete value in the abstract
world.
We also require the abstract domain to provide both sound forward
and backward operator for every syntactic operator of type
\fcode{binop} presented in Section~\ref{abint:section:lang-def}.
The function \fcode{abstract_binop} maps an operator \fcode{op} of
type \fcode{binop} to a sound forward abstract operator.
Its soundness is encoded as a refinement.
Similarly, \fcode{backward_abstract_binop} maps a \fcode{binop}
to a corresponding sound backward operator.
To ease backward analysis, \fcode{gt0} and \fcode{lt0} are
abstractions for non-null positive and negative integers.
\begin{fstarcode}
class num_adom (a: Type) = 
{ na_adom: adom a; adom_num: squash (na_adom.c == int_m)
; cgamma: xⵌ:a -> x:int_m -> b:bool {b <==> x ∈ gamma xⵌ}
; abstract_binop: op:_ -> i:a -> j:a -> r:a {lift op (gamma i) (gamma j) ⊆ gamma r}
; backward_abstract_binop: (op: binop) -> sound_backward_op a (concrete_binop op)
; gt0: xⵌ:a {forall(x:int_m). x>0 ==> x  ∈  gamma  xⵌ}
; lt0: xⵌ:a {forall(x:int_m). x<0 ==> x  ∈  gamma  xⵌ}; beta: x:int_m -> r:a{x  ∈  gamma  r} }
\end{fstarcode}
For a proposition \fcode{p}, the \gls{fstar} standard library
defines \fcode{squash p} as the type \fcode{_:unit{p}}, that is, a
refinement of the unit type. This can be seen as a lemma with no
parameter.
\subsubsection{Instance for intervals}
The section \ref{abint:section:intervals} defines everything required by
\fcode{num_adom}, thus below we give an instance of the typeclass
\fcode{num_adom} for intervals.
\begin{fstarcode}
instance itv_num_adom: num_adom itv = {
  na_adom = solve; adom_num = (); cgamma = itv_cgamma; beta = (λ x -> beta x);
  abstract_binop = (function | Plus -> itv_add ... | Or -> itv_ori);
  backward_abstract_binop = (function | Plus -> backward_add ... | Or -> backward_or );
  lt0 = (mk min_int_m (-1)); gt0 = (mk ( 1) max_int_m) }
\end{fstarcode}
\subsection{Memory Abstract Domains}
From the perspective of $\AbIntImp{}$ statements, an abstract domain
for abstract memories is fairly simple.
An abstract memory should be equipped with two operations:
assignment and assumption.
Those are directly related to their syntactic counterpart
\fcode{Assume} and \fcode{Assign}.
Thus, \fcode{mem_adom} has a field \fcode{assume_} and a field
\fcode{assign}.
The correctness of these operations are elegantly encoded as
refinement types.

Let us explain the refinement of \fcode{assume_}: let \fcode{mⵌ0}
an abstract memory, and \fcode{e} an expression.
For every concrete memory \fcode{m0} abstracted by \fcode{mⵌ0},
the set of acceptable final memories \fcode{osem_stmt (Assume e)
m0} should be abstracted by \fcode{assume_ mⵌ0 e}.
\begin{fstarcode}
class mem_adom 'm = { ma_adom: adom 'm; ma_mem: squash (ma_adom.c == mem);
  assume_: mⵌ0:'m -> e:expr -> mⵌ1:'m
    {forall (m0: mem{m0 ∈ gamma mⵌ0}). osem_stmt (Assume   e) m0 ⊆ gamma mⵌ1};
  assign: mⵌ0:'m -> v:varname -> e:expr -> mⵌ1:'m
    {forall (m0: mem{m0 ∈ gamma mⵌ0}). osem_stmt (Assign v e) m0 ⊆ gamma mⵌ1}}
\end{fstarcode}
\section{A Weakly-Relational Abstract Memory}
\label{abint:section:weakly-rel-amem}
In this section, we define a weakly-relational abstract memory.
This abstraction is said weakly-relational because the entrance of
an empty abstract value in the map systematically launches a
reduction of the whole map to \fcode{Bot}.
Below we define an abstract memory (\fcode{amem}) as either an
unreachable state (\fcode{Bot}), or a mapping (\fcode{map 't})
from \fcode{varname} to abstract values \fcode{'t}.
The mappings \fcode{map 'a} are equipped with the utility
functions \fcode{mapi}, \fcode{map1}, \fcode{map2} and
\fcode{fold}.
\begin{fstarcode}
type map 'a =!\shortdots!                 type amem 'a = withbot (map 't)
let get': map 'a -> varname -> 'a =!\shortdots!  let fold: ('a->'a->'a) -> map 'a -> 'a =!\shortdots!
let mapi: (varname -> 'a -> 'b) -> map 'a -> map 'b =!\shortdots!
let map1: ('a->'b) -> map 'a -> map 'b = fun f -> mapi (fun _ -> f)
let map2: ('a->'b->'c) -> map 'a -> map 'b -> map 'c =!\shortdots!
\end{fstarcode}
\subsubsection{A lattice structure}
The listing below presents \fcode{amem} instances for the
typeclasses \fcode{order}, \fcode{lattice} and \fcode{mem_adom}.
Once again, the various constraints imposed by these different
typeclasses are discharged automatically by the SMT solver.
\begin{fstarcode}
let amem_update (k: varname) (v: 't) (m: amem 't): amem 't
  = match m with | Bot -> Bot
    | Val m -> Val (mapi (fun k' v' -> if k'=k then v else v') m)
instance amem_lat {| l: adom 'a |}: lattice (amem 'a) =
  { corder = withbot_ord (fun m0 m1 -> fold (&&) (map2 corder  m0  m1))
  ; join = (fun x y -> match x, y with
      | Val x, Val y -> Val (map2 join x y) | m,Bot | _,m -> m)
  ; meet = (fun x y -> match x, y with
      | Val x, Val y ->
        let m = map2 (⊓) x y in
        if fold ( || ) (mapi (fun _ v -> l.adom_lat.corder v bottom) m)
        then Bot else Val m
      | _ -> Bot); bottom = Bot; top = ...}
instance amem_adom {|l:adom 'a|}: adom (amem 'a) = { c = mem' l.c
  ; adom_lat=solve; meet_law=(fun _ _->()); top_law=(fun _->()); bot_law=(fun _->())
  ; gamma = withbot_gamma (fun mⵌ m -> fold (/\) (mapi (fun v x -> m v ∈ gamma x) mⵌ))
  ; widen = (fun x y -> match x, y with
    | Val x, Val y -> Val (map2 widen x y) | m,Bot | _,m -> m)
  ; order_measure = let {max; f} = l.order_measure in
    { f = (function | Bot -> 0 | Val mⵌ -> 1 + fold (+) (map1 f mⵌ))
    ; max = 1 + max * 4 }}
\end{fstarcode}
The rest of this section defines a \fcode{mem_adom} instance for
our memories \fcode{amem}.
The typeclass \fcode{mem_adom} is an essential piece in our
abstract interpreter: it provides the abstract operations for
handling assumes and assignments.
\subsubsection{Forward expression analysis}
We define \fcode{asem_expr}, mapping expressions to abstract
values of type \fcode{'a}.
It is defined for any abstract domain, whence the typeclass
argument
\fcode{{|num_adom 'a|}}.
The abstract interpretation of an expression \fcode{e} given
\fcode{mⵌ0} an initial memory is defined below as
\fcode{asem_expr mⵌ0 e}.
It is specified via a refinement type to be a sound abstraction of
\fcode{e}'s operational semantics \fcode{osem_expr m0 e}.
This function leverages the operators from the different
typeclasses for which we defined instances just above.
\fcode{beta:int_m->'a} and \fcode{abstract_binop:binop->...} come
from \fcode{num_adom}, while \fcode{top:'a} comes from
\fcode{lattice}.
\begin{fstarcode}
val get: m:amem 'a {Val? m} -> varname -> 'a     let get (Val m) = get' m
let rec asem_expr {|num_adom 'a|} (mⵌ0: amem 'a) (e: expr)
 : (r: 'a { forall (m0: mem). m0 ∈ gamma mⵌ0 ==> osem_expr m0 e ⊆ gamma r })
 = if mⵌ0 ⊑ bottom then bottom else
   match e with | Const x -> beta x | Unknown -> top | Var v -> get mⵌ0 v
   | BinOp op x y -> abstract_binop  op  (asem_expr  mⵌ0  x) (asem_expr  mⵌ0  y)
\end{fstarcode}
\subsubsection{Backward analysis}
Our aim is to have an instance for our memory of \fcode{mem_adom}:
it expects an \fcode{assume_} operator. Thus, below a backward
analysis is defined for expressions.
Given an expression \fcode{e}, an abstract value \fcode{rⵌ} and a
memory \fcode{mⵌ0}, \fcode{backward_asem e rⵌ mⵌ} computes a new
abstract memory.
That abstract memory refines the abstract values held in
\fcode{mⵌ0} as much as possible under the hypothesis that \fcode{e}
lives in \fcode{rⵌ}.
The soundness of this analysis is encoded as a refinement on the
output memory. Given any concrete memory \fcode{m0} and integer
\fcode{v} approximated by \fcode{rⵌ}, if the operational semantics of
\fcode{e} at memory \fcode{m0} contains \fcode{v}, then \fcode{m0}
should also be approximated by the output memory.

When \fcode{e} is a constant which is not contained in the
concretization of the target abstract value \fcode{rⵌ}, the
hypothesis "\fcode{e} lives in \fcode{rⵌ}" is false, thus we
translate that fact by outputting the unreachable memory
\fcode{bottom}.
In opposition, when \fcode{e} is \fcode{Unknown}, the hypothesis
brings no new knowledge, thus we return the initial memory
\fcode{mⵌ0}.
In the case of a variable lookup (i.e. \fcode{e = Var v} for some
\fcode{v}), we consider \fcode{xⵌ}, the abstract value living at
\fcode{v}.
Since our goal is to craft the most precise memory such that
\fcode{Var v} is approximated by \fcode{rⵌ}, we alter \fcode{mⵌ0}
by assigning \fcode{xⵌ ⊓ rⵌ} at the variable \fcode{v}.
Finally, in the case of binary operations, we make use of the
backward operators and of recursion.
Note that it is the only place where we need to insert a hint for
the SMT solver: we assert an equality by asking \gls{fstar} to
normalize the terms.
We state explicitly that the operational semantics of a binary
operation reduces to two existentials: we manually unfold the
definition of \fcode{osem_expr} and \fcode{lift_binop}. The
\fcode{decreases} clause explains to \gls{fstar} why and how the
recursion terminates.
\begin{fstarcode}
let rec backward_asem {|l:num_adom 'a|} (e: expr) (rⵌ: 'a) (mⵌ0: amem 'a)
: Tot (mⵌ1: amem 'a { (* decreases: *) mⵌ1 ⊑ mⵌ0 /\ (* soundness: *)
      (forall(m0:mem) (v:int_m). (v ∈ gamma  rⵌ /\ m0 ∈ gamma  mⵌ0 /\ v ∈ osem_expr m0 e)
                               ==> m0 ∈ gamma mⵌ1)}) (decreases e)
  = if mⵌ0 ⊑ bottom then bottom else match e with
  | Const x -> if cgamma rⵌ x then mⵌ0 else bottom | Unknown -> mⵌ0
  | Var v -> let xⵌ: 'a =  rⵌ ⊓ get  mⵌ0 v  in
            if xⵌ ⊑ bottom then Bot else amem_update v xⵌ  mⵌ0
  | BinOp op ex ey -> let backward_op = backward_abstract_binop op in
      let xⵌ, yⵌ = backward_op (asem_expr mⵌ0 ex) (asem_expr mⵌ0 ey) rⵌ in
      let rⵌ: amem 'a = backward_asem ex xⵌ mⵌ0 ⊓ backward_asem ey yⵌ mⵌ0 in
      assert_norm (forall (m: mem) (v: int_m). v ∈ osem_expr m e
        <==> (exists (x y:int_m). x ∈ osem_expr m ex /\ y ∈ osem_expr m ey
                         /\ v == concrete_binop op x y));
      rⵌ
\end{fstarcode}
\subsubsection{Iterating the backward analysis}
While a concrete test is idempotent, it is not the case for
abstract ones.
Our goal is to refine an abstract memory under a hypothesis as
much as possible.
Since \fcode{backward_asem} is proven sound and decreasing, we can
repeat the analysis as much as we want. We introduce
\fcode{prefixpoint} that computes a pre-fixpoint.
However, even if the function from which we want to get a
prefixpoint is decreasing, this is not a guarantee for
termination.
The type \fcode{measure} below associates an order to a measure
that ensures termination.
Such a measure cannot be implemented for a lattice that has
infinite decreasing or increasing chains.
We also require a maximum for this measure, so that we can reverse
the measure easily in the context of postfixpoints iteration.
\begin{fstarcode}
type measure #a (ord: a -> a -> bool) 
  = { f: f: (a -> nat) {forall x y. x `ord` y ==> x !\notPropEq{}! y ==> f x < f y}
    ; max: (max: nat {forall x. f x < max}) }
\end{fstarcode}
Let us focus on \fcode{prefixpoint}: given an order \fcode{⊑} with
its measure \fcode{m}, it iterates a decreasing function \fcode{f},
starting from a value \fcode{x}.
The argument \fcode{r} is a binary relation which is required to
hold for every couple \fcode{(x, f x)}.
\fcode{r} is also required to be transitive, so that morally
\fcode{r x (fⁿ x)} holds.
\fcode{prefixpoint} is specified to return a prefixpoint
\fcode{y}, that is, with \fcode{r x y} holding.
\begin{fstarcode}
let rec prefixpoint ((⊑): order 'a) (m: measure (⊑))
  (r: 'a->'a->prop {trans r}) (f: 'a->'a {∀e. f e ⊑ e /\ r e (f e)}) (x:'a) 
  : Tot (y: 'a{r x y /\ f y == y /\ y ⊑ x}) (decreases (m.f x))
  = let x' = f x in if x ⊑ x' then x else prefixpoint (⊑) m r f x'
\end{fstarcode}
Below is defined \fcode{backward_asem_fp} the iterated version of
\fcode{backward_asem}.
Besides using \fcode{prefixpoint}, the only thing required here is
to spell out \fcode{t}, the relation we want to ensure.
\begin{fstarcode}
let backward_asem_fp {|num_adom 'a|} (e:expr) (r:'a) (mⵌ0:amem 'a)
  : Tot (mⵌ1: amem 'a {(forall (m0:mem) (v:int_m). mⵌ1 ⊑ mⵌ0 /\
                   (v ∈ gamma r /\ m0 ∈ gamma mⵌ0 /\ v ∈ osem_expr m0 e) ==> m0 ∈ gamma mⵌ1)})
  = let t (mⵌ0 mⵌ1: amem 'a) = forall (m: mem) (v: int_m).
      (v ∈ gamma r /\ m ∈ gamma mⵌ0 /\ v ∈ osem_expr m e) ==> m ∈ gamma mⵌ1 in
    prefixpoint corder order_measure t (backward_asem e r) mⵌ0
\end{fstarcode}
\subsubsection{A \fcode{mem_adom} instance} We defined both a
forward and backward analysis for expressions. Implementing an
\fcode{mem_adom} instance for \fcode{amem} is thus easy, as shown
below.
For any numerical abstract domain \fcode{'a},
\fcode{amemory_mem_adom} provides an \fcode{mem_adom}, that is,
an abstract domain for memories, providing nontrivial proofs of
correctness. Still, this is proven automatically.
\begin{fstarcode}
instance amemory_mem_adom {| nd: num_adom 'a |}: mem_adom (amem 'a) =
  let adom: adom (amem 'a) = amem_adom in { ma_adom = adom; ma_mem = ()
  ; assume_ = (fun mⵌ e -> backward_asem_fp e gt0 mⵌ ⊔ backward_asem_fp e lt0 mⵌ)
  ; assign = (fun mⵌ v e -> let vⵌ: 'a = asem_expr mⵌ e in
                if vⵌ ⊑ bottom then Bot else amem_update v vⵌ mⵌ)}
\end{fstarcode}
\section{Statement Abstract Interpretation}
\label{abint:section:stmt-ab-int}
Wrapping up the implementation of our abstract interpreter, this
section presents the abstract interpretation of $\AbIntImp{}$
statements.
For every memory type \fcode{'m} that instantiates the typeclass
of abstract memories \fcode{mem_adom}, the abstract semantics
\fcode{asem_stmt} maps statements and initial abstract memories
to final memories.
\fcode{mem_adom} is defined and proven correct below.

Given a statement \fcode{s}, and an initial abstract memory
\fcode{mⵌ0}, \fcode{mem_adom s mⵌ0} is a final abstract memory so
that for any initial concrete memory \fcode{m} approximated by
\fcode{mⵌ0} and for any acceptable final concrete memory
\fcode{m'} considering the operational semantics, \fcode{m'} is
approximated by \fcode{mem_adom s mⵌ0}.
Here, we give two hints to the SMT solver: by normalization
(\fcode{assert_norm}), we unfold the operational semantics in the
case of choices or sequences.
The analysis of an assignment or an assume is very easy since we
already have operators defined for these cases.
In the case of the sequence of two statements, we simply recurse.
Similarly, when the statement is a choice, we recurse on its two
possibilities.
Then the two resulting abstract memories are merged back together.
The last case to be handled is the loop, that is some statement of
the shape \fcode{Loop body}.
We compute a fixpoint \fcode{mⵌ1} for \fcode{body}, by widening:
it therefore approximates correctly the operational semantics of
\fcode{Loop body}, since it is defined as a transitive closure.
\gls{fstar}'s standard library provides the lemma
\fcode{stable_on_closure}; of which we give a simplified signature
below. The concretization \fcode{gamma mⵌ1} is a set, that is a
predicate: we use this lemma with \fcode{gamma mⵌ1} as predicate
\fcode{p} and with the operational semantics as relation
\fcode{r}.
\begin{fstarcode}
val simplified_stable_on_closure: r:('a -> 'a -> prop) -> p:('a -> prop)
  -> Lemma (requires forall x y. p x /\ r x y ==> p y)
          (ensures forall x y. p x /\ closure r x y ==> p y)
\end{fstarcode}

\begin{fstarcode}
let rec asem_stmt {| md: mem_adom 'm |} (s: stmt) (mⵌ0: 'm)
  : (mⵌ1:'m {forall(m m':mem). (m ∈ gamma  mⵌ0 /\ m' ∈ osem_stmt s  m) ==> m' ∈ gamma  mⵌ1})
  = assert_norm(∀s0 s1 (m0 mf:mem). osem_stmt (Seq s0 s1) m0 mf
     == (exists(m1:mem). m1 ∈ osem_stmt s0 m0 /\ mf ∈ osem_stmt s1 m1));
    assert_norm(∀a b (m0 mf:mem). osem_stmt (Choice a b) m0 mf
     == (mf ∈ (osem_stmt a m0 ∪ osem_stmt b m0)));
    if mⵌ0 ⊑ bottom then bottom
    else match s with         | Assign v e -> assign mⵌ0 v e
    | Assume e -> assume_ mⵌ0 e | Seq s t -> asem_stmt t (asem_stmt s mⵌ0)
    | Choice a b -> asem_stmt a mⵌ0 ⊔ asem_stmt b mⵌ0
    | Loop body -> let mⵌ1: 'm = postfixpoint corder order_measure
                   (fun (mⵌ:'m) -> widen mⵌ (asem_stmt body mⵌ <: 'm))
                  in stable_on_closure (osem_stmt body) (gamma mⵌ1) (); mⵌ1
\end{fstarcode}
Below we show the definition of \fcode{postfixpoint}, which is
similar to \fcode{prefixpoint}. However, it is simpler because
it only ensures its outcome is a postfixpoint.
\begin{fstarcode}
let rec postfixpoint ((⊑): order 'a) (m: measure (⊑))
  (f: 'a -> 'a {forall x. x ⊑ f x}) (x: 'a)
  : Tot (y: 'a{f y == y /\ (⊑) x y}) (decreases (m.max - m.f x))
  = let x' = f x in if x' ⊑ x then x else postfixpoint (⊑) m f x'
\end{fstarcode}

\section{Conclusion and further works}%
\label{abint:section:conclusion}%

We presented almost the entire code of our abstract interpreter for $\AbIntImp{}$.
Our approach to abstract interpretation is concretization-based, and
follows the methodology of~\cite{SAS13:Blazy:al,POPL15:Jourdan:al}.
While using \gls{fstar}, we did not encountered any issue regarding
expressiveness, and additionally gained a lot in proof automatization,
to finally implement a fairly modular abstract interpreter.
The table below compares the line-of-proof vs. line-of-code ratio of
our implementation compared to some of the available verified abstract
interpreters.
Ours is up to 17 times more proof efficient.
It is very compact, and requires a negligible amount of manual proofs.
This comparison has its limits, since the different formalizations do
not target the same programming languages:~\cite{POPL15:Jourdan:al}
and~\cite{SAS13:Blazy:al} handles the full C language,
while~\cite{ITP10:Cachera:Pichardie} and the curent paper deal with
more simple imperative languages.
Also, proof effort usually does not scale linearly.

\begin{table}
  \centering%
  \setlength{\tabcolsep}{2mm}
  \begin{tabular}{rrrcl}
     & Code & Proof & Ratio & Feature set\\
    \hline
    This paper
    & 487   & 39    & 0.08 & Simple imperative language
    \\
    Pichardie et al.~\cite{ITP10:Cachera:Pichardie}
    & 3\,725  & 5\,020  & 1.35 & Simple imperative language
    \\
    Verasco~\cite{POPL15:Jourdan:al}
    & 16\,847 & 17\,040 & 1.01 & CompCert C langage
    \\
    Blazy et al.~\cite{SAS13:Blazy:al}
    & 4\,000  & 3\,500  & 0.87 & CompCert C langage
  \end{tabular}%
\end{table}

The sources of our abstract interpreter sources are available along
with a set of example programs; building it natively or as a web
application is easy, reproducible\footnote{
  Our build process relies on the purely functional Nix package manager.%
} and automated.

This work is very far from the scope of Verasco which required about
four years of human
time~\cite{laporte:tel-01285624,jourdan:tel-01327023}, but our
results, which required 3 months of work with \gls{fstar} expertise,
are very encouraging.

\paragraph{Further work} We aim at following the path of Verasco by adding
real-world features to our abstract interpreter and consider a more
realistic target language such as one of the CompCert C-like input
languages.
One of the weakenesses of Verasco is its efficiency. Using
\gls{lowstar}, a C DSL for \gls{fstar}, it is possible to write (with
a nontrivial additionnal effort related to \gls{lowstar}) a very
efficient C and formally verified abstract interpreter.
This development also opens the path for enriching \gls{fstar}
automation via verified abstract interpretation.

\section*{Acknowledgements}
This work is supported by a European Research Council (ERC)
Consolidator Grant for the project VESTA, funded under the
European Union’s Horizon 2020 Framework Programme (grant agreement
772568).
\newpage
\printglossary[type=\acronymtype]
\bibliographystyle{latex/splncs04}
\bibliography{latex/main}

\end{document}